\documentclass[10pt,a4paper]{article} 

\usepackage{hyperref}
\hypersetup{
  colorlinks   = true, 
  urlcolor     = blue, 
  linkcolor    = blue, 
  citecolor   = red 
}

\usepackage{amsmath}
\usepackage{amsthm}
\usepackage{amsfonts}
\usepackage{amssymb}
\usepackage{array}
\usepackage[top=3.5cm,bottom=3.5cm,left=4cm,right=4cm]{geometry} 
\usepackage{authblk}
\usepackage{color, colortbl}
\usepackage{enumitem}
\usepackage{numprint}
\npthousandsep{\,}

\definecolor{Gray}{gray}{0.2}
\newtheorem{theorem}{Theorem}[section]

\newtheorem{proposition}[theorem]{Proposition}
\newtheorem{corollary}[theorem]{Corollary}
\newtheorem{lemma}[theorem]{Lemma}

\theoremstyle{definition}
\newtheorem{definition}[theorem]{Definition}

\newtheorem{remark}[theorem]{Remark}

\newcommand{\C}{\mathcal{C}}

\newcommand{\F}{\mathbb{F}}

\newcommand{\N}{\mathbb{N}}

\newcommand{\Sn}{\mathcal S_n}

\newcommand{\Fq}{\mathbb{F}_{q}}

\newcommand{\Z}{\mathbb{Z}}

\usepackage[dvipsnames]{xcolor}
\usepackage{pagecolor}
\definecolor{light-gray}{gray}{0.90}

\usepackage{tikz}
\usepackage{pgfplots}

\title{\textbf{New Lower Bounds for Permutation Codes using Linear Block Codes}}

\author[1]{Giacomo Micheli \thanks{The author was supported by the Swiss National Science Foundation through grant no. 171249.}}
\author[2]{Alessandro Neri\thanks{The author was supported by the Swiss National Science Foundation through grant no. 169510.}}

\affil[1]{Institute of Mathematics, EPFL, Switzerland}
\affil[2]{Institute of Mathematics, University of Zurich, Switzerland}

\setlist[itemize]{noitemsep, nolistsep}

\date{}

\begin{document}
\maketitle
\thispagestyle{empty}

\begin{abstract}
In this paper we prove new lower bounds for the maximal size of permutation codes by connecting the theory of permutation codes with the theory of linear block codes.
More specifically, using the columns of a parity check matrix of an $[n,k,d]_q$ linear block code, we are able to prove the existence of a permutation code in the symmetric group of degree $n$, having minimum distance at least $d$ and large cardinality.
With our technique, we obtain new lower bounds for permutation codes that enhance the ones in the literature and provide  asymptotic improvements in certain regimes of length and distance of the permutation code.
\end{abstract}

\section{Introduction}

Permutation codes have been of great interest recently due to their applications (for example in powerline communications \cite{ch04,co04}) and for their intrinsecal combinatorial interest \cite{fr77,ga13,ji16,ta99,wa17}. Let us now briefly explain what permutation codes are.
The symmetric group $\mathcal S_n$ can be endowed with a metric  $d_h$ defined as follows: if $\sigma,\tau\in \mathcal S_n$, then $d_h(\sigma,\tau)=|\{i\in\{1,\dots, n\}: \: \sigma(i)\neq \tau(i)\}|$. An $(n,d)$-permutation code is a subset $\Gamma$ of $\mathcal S_n$ such that $\min\{d_h(\sigma,\tau): \: \sigma,\tau\in \Gamma, \sigma \neq \tau \}=d$.
The maximal size $M(n,d)$ of an  $(n,d)$-permutation code has been studied widely in the literature. Very nice ideas to produce lower bounds appeared in \cite{ga13,ji16,wa17}, and they all improve asymptotically the famous Gilbert-Varshamov bound.  In this paper we provide new lower bounds for $M(n,d)$. From a theoretical point of view, the paper connects the theory of permutation codes with the theory of linear block codes and converts the problem of extistence of permutation codes with certain parameters into existence problems for some linear block codes. From a practical perspective, our approach allows to produce improved bounds for many set of parameters $n,d$. Moreover, for certain choices of regimes of $n$ and $d$ we actually beat asymptotically the best known bounds in \cite{ji16,wa17}. 
The paper is structured as it follows. 

Section \ref{sec:prelim} recaps the basic tools we need from coding theory and the theory of permutation codes. 

Section \ref{sec:lowerbounds} provides the technical heart of our proof, which gives the wanted connection  between the theory of permutation codes and the theory of linear block codes. 

In Section \ref{sec:lowerusingMDS} we use the results of Section \ref{sec:lowerbounds} together with results from the theory of Maximum Distance Separable (MDS) codes to provide two new lower bounds on permutation codes. The first (Theorem \ref{cor:MDSq}) beats the bounds in \cite{ji16,wa17} whenever  the next prime power larger than or equal to $n$ is smaller than  the next prime larger than or equal to $n$ (in all the other cases it gives the same bound). The second one (Theorem \ref{thm:boundMDSq+1}) beats asymptotically \cite{ji16,wa17} in the large distance regime.

In Section \ref{sec:lowerboundsusingAMDS} we produce new bounds using Almost MDS codes that provide additional improvements of the bounds in \cite{ji16, wa17} under the assumption that a linear code with certain parameters exists.

Finally, in Section \ref{sec:comparison} we compare the bounds we obtained in the paper with the previous bounds in the literature.

Conclusions are provided in Section \ref{sec:conclusions}.

\section{Preliminaries}\label{sec:prelim}
In this section we recall the basic notions of linear codes endowed with the Hamming distance, and the theory of permutation codes. 
\subsection{Linear Block codes}
Let $q$ be a prime power and denote by $\Fq$ the field with $q$ elements. For a given positive integer $n$ we consider, the \textbf{Hamming distance}  over $\Fq^n$, that is the map
$$d_H:\Fq^n \times \Fq^n \longrightarrow \N,$$
 defined by $d_H(u,v)=|\{i \in \{1,\ldots, n\} \mid u_i \neq v_i\}|$ for $u=(u_1,\ldots, u_n), v=(v_1,\ldots, v_n) \in \Fq^n$.
Moreover, the \textbf{Hamming weight} of a vector $v \in \Fq^n$ is the quantity \[w_H(v)=d_H(v,0)=|\{i \in \{1,\ldots,n\}\mid v_i \neq 0\}|\].

In this context, an \textbf{$[n,k]_q$  code} $\C$ is a $k$-dimensional subspace of $\Fq^n$ equipped with the Hamming distance. The integer $n$ is the \textbf{length} and $k$ is called the \textbf{dimension} of  $\C$.  The \textbf{minimum distance} of  $\C$ is the integer defined by
$$d(\C):=\min\{d_H(u,v) \mid u,v \in \C, u \neq v \}.$$
In the following we will use the notation $[n,k,d]_q$  for a code of length $n$, dimension $k$ and minimum distance $d$.

\begin{definition}
The \textbf{dual code} $\C^\perp$ of an $[n,k]_q$ code $\C$ is the $[n,n-k]_q$ code
$$\C^\perp:=\left\{u \in \Fq^n \mid \langle u,c \rangle=0 \mbox{ for all } c \in \C\right\},$$
where $\langle \cdot , \cdot\rangle$ denotes the standard inner product between two vectors in $\Fq^n$.
\end{definition}

Two important matrices are related to an $[n,k]_q$ code $\C$. A \textbf{generator matrix} $G\in \Fq^{k\times n}$ for $\C$ is a $k \times n$ matrix in $\Fq$ whose rows are a basis for $\C$, i.e. $\C=\{mG \mid m \in \Fq^k\}$. A \textbf{parity check matrix} for $\C$ is a matrix $H \in \Fq^{(n-k) \times n}$ such that $\C=\{u \in \Fq^n \mid Hu^\top=0\}$.

From the definition, it is straightforward to verify that a matrix $H \in \Fq^{(n-k) \times n}$ is a parity check matrix for an $[n,k]_q$ code $\C$ if and only if it is a generator matrix for the dual code $\C^\perp$.
%

\begin{proposition}
Let $\C$ be an $[n,k]_q$ code,  $H\in \Fq^{(n-k)\times n}$ be a parity check matrix for $\C$ and let $d$ be a positive integer. The following are equivalent.
\begin{enumerate}
\item $d(\C) \geq d$.
\item Every $d-1$ columns of $H$ are linearly independent over $\Fq$.
\end{enumerate}
\end{proposition}

\begin{definition}
 Two $[n,k]_q$ codes $\C$ and $\C'$ are said to be \textbf{equivalent} if there exists $\sigma \in \Sn$, $\lambda_1, \ldots, \lambda_n \in \Fq^*$ such that
$$\C'=\left\{ (\lambda_1c_{\sigma(1)}, \ldots, \lambda_nc_{\sigma(n)}) \mid (c_1,\ldots, c_n) \in \C\right\}.$$
\end{definition}

In terms of their generator matrices an, respectively, parity check matrices, we can see the following. If $G$ and $G'$ are  generator matrices for $\C$ and $\C'$ respectively, then $\C$ and $\C'$ are equivalent if and only if there exists $P$ permutation matrix and $D$ diagonal matrix such that $G'=GPD$. An analogous statement holds with their parity check matrices. 

\begin{proposition}
Let $H$ and $H'$ be  parity check matrices for two $[n,k]_q$ codes $\C$ and $\C'$ respectively. Then, $\C$ and $\C'$ are equivalent if and only if there exists a  permutation matrix $P$ and a diagonal matrix $D$ such that $H'=HPD$.
\end{proposition}

\begin{lemma}\label{lem:firstrow1}
 Let $\C$ be an $[n,k]$ linear code $\C$. If $\C^\perp$ has a codeword of Hamming weight $n$, then there exists an $[n,k]$ code $\C'$  equivalent to $\C$ which has a parity check matrix whose first row is equal to $(1,1,\ldots,1)$.
\end{lemma}

\begin{proof}
 A parity check matrix for $\C$ is a generator matrix for $\C^\perp$.  Let $v \in \C^\perp$ be a codeword of Hamming weight $n$, and take as a generator matrix for $\C^\perp$ a matrix $H$ whose first row is $v=(v_1,\ldots, v_n)$. Define the matrix $D=\mathrm{diag}(v_1^{-1},\ldots, v_n^{-1})$. Therefore, the code $\C'$ whose parity check matrix is $H'=HD$ is  equivalent to $\C$ and the first row of $H'$ is equal to $(1,1,\ldots,1)$.
\end{proof}

\subsection{Permutation codes}

Let $n\in \N$ be a positive integer and denote by $\Sn$ the symmetric group on $n$ elements.  
On the group $\Sn$ we consider the  \textbf{Hamming distance}, that is defined for $\sigma, \tau \in \Sn$, as
$$d_h(\sigma, \tau)=|\left\{i \in \{1,\ldots,n\} \mid \sigma(i)\neq \tau(i) \right\}|.$$

\begin{definition}
A \textbf{permutation code} of length $n$ is a subset $\Gamma$ of $\Sn$ endowed with the Hamming distance. The minimum distance of $\Gamma$ is the quantity
$$d(\Gamma)=\min\{d_h(\sigma, \tau) \mid \sigma,\tau \in \Gamma, \sigma \neq \tau\}.$$
\end{definition}

Let $M(n,d)$ be the maximum cardinality that a permutation code of length $n$ and minimum distance $d$ can have. There are many known bounds on this quantity, that we now briefly recall.

\begin{theorem}[Singleton-like bound]
$$M(n,d) \leq \frac{n!}{(d-1)!}.$$
\end{theorem}

A \textbf{derangement} of size $r$ is a permutation on $r$ elements with no fixed points. Let $D_r$ denote the number of derangements of size $r$. The number of derangements of size $k$ is also known as the \emph{subfactorial}, and it is well-known that
$$D_r=r!\sum_{i=0}^r\frac{(-1)^i}{i!}=\left\lfloor\frac{r!}{e}+\frac{1}{2} \right\rfloor.$$

\begin{theorem}[Sphere-packing bound]
$$M(n,d)\leq \frac{n!}{\sum_{i=0}^{\lfloor\frac{d-1}{2}\rfloor} \binom{n}{i}D_i}.$$
\end{theorem}

\begin{theorem}[Gilbert-Varshamov bound]
$$M(n,d)\geq \frac{n!}{\sum_{i=0}^{d-1} \binom{n}{i}D_i}.$$
\end{theorem}

An improvement of the Gilbert-Varshamov bound, at least from an asimptotical point of view, was given in \cite{ji16}, whose proof relies on rational function fields theory. Another proof of the same result can be found  in \cite{wa17}.

\begin{theorem}\cite[Theorem 2]{ji16}\cite[Theorem 13]{wa17}.
 For every prime $p\geq n$, for every $2<d\leq n$,
$$M(n,d) \geq \frac{n!}{p^{d-2}}.$$
\end{theorem}

\section{Bounding Permutation Codes Using Linear Block Codes}\label{sec:lowerbounds}

In this section we provide a general lower bound on the maximal size of a permutation code of given length $n$ and minimum distance $d$. The bound in Theorem \ref{thm:firstbound} is the technical heart of the paper from which the explicit bounds in the next sections will follow.

Let $n$ be a positive integer. For a given subset $\mathcal K$ of the symmetric group $\Sn$, we denote by $M(\mathcal K, d)$ the maximum cardinality of a permutation code of minimum distance at least $d$ entirely contained in $\mathcal K$, i.e.
$$M(\mathcal K, d)=\max\left\{ |\Gamma| \mid  \Gamma \subseteq \mathcal K, d(\Gamma)\geq d \right\}.$$
Note that, with this notation,  $M(\Sn,d)=M(n,d)$.
In the next proposition we use the convention that $\mathcal S_0=\mathcal S_1=\{1\}$. For a set $A\subset \mathcal S_n$ and an element $g\in \mathcal S_n$ we denote by $Ag$ the set $\{ag:\: a\in A\}$. Clearly, if $\Gamma$ is a permutation code of minimum distance $d$, then also $\Gamma g$ is a permutation code of minimum distance $d$.
\begin{theorem}\label{thm:firstbound}
Let $d,k,n$ be integers such that $0<k<n$ and $1<d\leq n$. Let moreover $q$ be a prime power and $s,r$ be positive integers such that $n=qs+r$ and $0 \leq r <q$. If there exists  an $[n,k,d]_q$ code $\C$ such that $\C^\perp$ has a codeword of Hamming weight $n$, then
$$M(n,d) \geq \frac{n!M(\mathcal K,d)}{(s+1)!^rs!^{q-r}q^{n-k-1}},$$
where $\mathcal K=(\mathcal S_{s+1})^{r}\times (\mathcal S_{s})^{q-r}$.
\end{theorem}

\begin{proof}
Let $\C$ be an $[n,k,d]_q$ code such that $\C^\perp$ has a codeword of Hamming weight $n$. By Lemma \ref{lem:firstrow1} we have an $[n,k,d]_q$ code $\C'$ with a parity check matrix $H\in \Fq^{(n-k)\times n}$  whose first row is $(1,1,\ldots, 1)$. Let $v_i$ be the $i$-th column of $H$ and let $n=qs+r$ with $0\leq r <q$. We can write 
$\Fq=\left\{a_0,\ldots, a_{q-1}\right\}$ and define the map
$$L:\{1,\ldots,n\} \longrightarrow \Fq: i \longmapsto a_{(i\, \mathrm{mod} \,q)}.$$
Moreover, choose the subgroup of $\Sn$ defined as 
$$\mathcal K=\left\{\sigma \in \Sn \mid \sigma(i)\equiv i \mod q, \mbox{ for all } i\in\{1,\ldots, n\} \right\}.$$
One can see that $\mathcal K\cong (\mathcal S_{s+1})^r \times (\mathcal S_{s})^{q-r}$. 

Let $\Gamma' \subseteq \mathcal K$ be a permutation code of minimum distance $d$ and cardinality $M(\mathcal K, d)$. Consider the set of right cosets of $\Sn/\mathcal K$, that is $\{\mathcal K\sigma_i \}_{i\in \{1,\dots, |\Sn|/|\mathcal K|\}}$  for some $\sigma_i$'s in $\mathcal S_n$. Define the set 
$$\mathcal T:=\bigcup_i  \Gamma'\sigma_i .$$
From this set, we consider the map
$$\begin{array}{rcl}
 \varphi:\mathcal T & \longrightarrow & \Fq^{n-k}\\
\sigma & \longmapsto & \sum\limits_{i=1}^{n} L(\sigma(i))v_i.
\end{array}$$

 Assume $\varphi(\sigma)=\varphi(\tau)$ and $d_H(\sigma,\tau)=r\leq d-1$. Let $\{j_1,\ldots,j_r\}$ be the subset of $\{1,\ldots, n\}$ such that $\sigma(j_i) \neq \tau(j_i)$. Then
$$0=\varphi(\sigma)-\varphi(\tau)=\sum_{\ell=1}^r(L(\sigma(j_\ell))-L(\tau(j_\ell)))v_{j_\ell}.$$
Since $v_{j_1}, \ldots, v_{j_r}$ are linearly independent, it follows $L(\sigma(j_\ell))-L(\tau(j_\ell))=0$ for every $\ell \in \{1,\dots, r\}$. 
 Therefore, $\sigma$ and $\tau$ are equal over the integers on all the $i$'s  not in $\{j_1,\dots j_\ell\}$ (because of their distance), and they are equal modulo $q$ on all the $i$'s in $\{j_1,\dots j_\ell\}$ (since the $a_i$ are all distinct elements of $\mathbb F_q$ and by the independence of the $v_{j_\ell}$'s). This forces in particular that $\sigma(i)\equiv \tau(i)\mod q$ for any $i\in \{1,\dots n\}$. Since the equation holds for any $i$, by relabeling $i$ with $\tau(i)$, we get that $\sigma\tau^{-1}(i) \equiv i \mod q$ for all $i\in \{1,\dots n\}$. This implies that $\sigma\tau^{-1}\in \mathcal K$ and also, by construction, we have $\Gamma'\sigma = \Gamma'\tau$. Since $d_H(\sigma,\tau)<d$ and $d(\Gamma'\sigma)=d(\Gamma')=d$, we obtain $\sigma=\tau$. This shows that for every $z\in \mathrm{Im}(\varphi)$ the preimage $\varphi^{-1}(z)$ is a permutation code of minimum distance at least $d$. Moreover, since $H$ has $(1,1,\dots,1)$ as first row, $ \mathrm{Im}(\varphi) \subseteq \mathcal H_1$, where 
$$\mathcal H_1=\{(x_1,\ldots, x_{n-k}) \in \Fq^{n-k} \mid x_1=\sum_{i=1}^{n} L(i)\}.$$
Therefore, by generalized pigeonhole principle, we have that there exists $z\in \mathcal H_1$ such that $\varphi^{-1}(z)$ has cardinality at least
$$ \frac{|\mathcal T|}{|\mathrm{Im}(\varphi)|}\geq \frac{|\mathcal T|}{|\mathcal H_1|}= \frac{n!M(\mathcal K,d)}{(s+1)!^rs!^{q-r}q^{n-k-1}}.$$

\end{proof}

In the rest of the paper we will apply Theorem \ref{thm:firstbound}, as we will be always able to show the existence of a codeword of weight $n$ in the dual of the code. Nevertheless, one can also show the following
\begin{theorem}\label{thm:firstbound2}
Let $d,k,n$ be integers such that $0<k<n$ and $1<d\leq n$. Let moreover $q$ be a prime power and $s,r$ be positive integers such that $n=qs+r$ and $0 \leq r <q$. If there exists  an $[n,k,d]_q$ code $\C$, then we have
$$M(n,d) \geq \frac{n!M(\mathcal K,d)}{(s+1)!^rs!^{q-r}q^{n-k}},$$
where $\mathcal K=(\mathcal S_{s+1})^{r}\times (\mathcal S_{s})^{q-r}$.
\end{theorem}
\begin{proof}
The proof is completely analogous except for the fact that $ \mathrm{Im}(\varphi)$ is not anymore included in $\mathcal H_1$ (as $H$ does not necessarily has in the first row all $1$'s). Therefore, in the last step one simply has to replace $\mathcal H_1$ with $\mathbb F_q^{n-k}$ getting 
$$ \frac{|\mathcal T|}{|\mathrm{Im}(\varphi)|}\geq \frac{|\mathcal T|}{|\mathbb F_q^{n-k}|}= \frac{n!M(\mathcal K,d)}{(s+1)!^rs!^{q-r}q^{n-k}}.$$
\end{proof}

\section{Lower bounds using MDS codes}\label{sec:lowerusingMDS}

In this section we are going to apply the result of Theorem \ref{thm:firstbound} using a specific class of linear codes, namely the MDS codes.

\begin{theorem}[Singleton Bound \cite{si64}]\label{thm:SingBound} 
 Let $\C$ be an $[n,k,d]_q$ code. Then
$$d \leq n-k+1.$$
\end{theorem}

The \textbf{Singleton defect} of an $[n,k,d]_q$ code $\C$  is the number $s(\C):=n-k+1-d$. Observe that, by Theorem \ref{thm:SingBound}, the Singleton defect of a linear code $\C$ is always a non-negative integer.

Recall that, for fixed $n$ and $d$, the lower bound on $M(n,d)$ provided in Theorem \ref{thm:firstbound} depends on the existence of an $[n,k,d]_q$ code $\C$, and it contains a factor $q^{n-k-1}$ in the denominator. Since $n-k-1=d-2+s(\C)$, it is only useful to consider codes with small Singleton defect.

\begin{definition}
An $[n,k,d]_q$ code $\C$ with $s(\C)=0$  is called \textbf{maximum distance separable (MDS) code}.
\end{definition}
Whenever an $[n,k,d]_q$-code is MDS, we write that is an $[n,k]_q$ MDS code.

MDS codes have been deeply studied over the last 60 years because of their optimal parameters \cite{ma77, va12} and their connection to finite projective geometry \cite{se55, br88}. In the following we recall few of their basic properties.

\begin{theorem}\label{thm:dualMDS}
Let $\C$ be an $[n,k]_q$ MDS code. Then $\C^\perp$ is an $[n.n-k]_q$ MDS code.
\end{theorem}

\begin{theorem}\cite[Theorem 6]{ez11}\label{thm:qMDSweights}
Any $[n, k]_q$ MDS code with $n \leq q$ has a codeword of weight $\ell$ for every $\ell=n-k+1,\ldots,n$. In particular, for every $k$, a $[q,k]_q$ code has codewords of weight $q$.
\end{theorem}

\begin{corollary}\label{cor:MDSq}
For every $k$ and every $[q,k]_q$ MDS code $\C$, the dual code $\C^\perp$ has a codeword of weight $q$.
\end{corollary}

\begin{proof}
Let $\C$ be a $[q,k]_q$ MDS code. By Theorem \ref{thm:dualMDS}, $\C^\perp$ is a $[q,q-k]_q$ MDS code, and by Theorem \ref{thm:qMDSweights}, $\C^\perp$ has a codeword of Hamming weight $q$.
\end{proof}

\begin{theorem}\label{thm:boundMDSq}
For every prime power $q \geq n$, and every integer $d$ with $2<d<n$,
$$M(n,d) \geq \frac{n!}{q^{d-2}}.$$
\end{theorem}

\begin{proof}
It directly follows from Theorem \ref{thm:firstbound}  with the choice, $s=1$, and $r=0$, and Corollary \ref{cor:MDSq}  which ensures the existence of the wanted $[n,k,d]_q$-code.
\end{proof}

Theorem \ref{thm:boundMDSq} provides a lower bound on $M(n,d)$, using the existence of MDS codes of length $n$ over a finite field with cardinality at least $n$. The rest of the section is devoted to obtain a similar bound, using   MDS codes whose length exceeds the cardinality of the underlying  finite field.

\begin{theorem}\cite[Theorem 8]{ez11}
A $[q+1, k]_q$ MDS code has a codeword of weight $\ell$ for every $\ell\in \{q-k+2,\ldots,q+1\}$, except for the $q$-ary symplex code $[q+1,2]_q$, that has only codewords of weight $0$ and $q$. In particular, for every $k\neq 2$, a $[q+1,k]_q$ code has codewords of weight $q+1$.
\end{theorem}

\begin{corollary}\label{cor:MDSq+1}
For every $k \neq q-1$ and every $[q+1,k]_q$ MDS code $\C$, the dual code $\C^\perp$ has a codeword of weight $q+1$.
\end{corollary}

\begin{proof}
Let $\C$ be a $[q+1,k]_q$ MDS code. By Theorem \ref{thm:dualMDS}, $\C^\perp$ is a $[q+1,q+1-k]_q$ MDS code, with $q+1-k \neq 2$. Therefore, by Theorem \ref{thm:qMDSweights}, $\C^\perp$ has a codeword of Hamming weight $q+1$.
\end{proof}

\begin{theorem}\label{thm:boundMDSq+1}
For every prime power $q$, and every $3<d<q$,
$$M(q+1,d) \geq \frac{(q+1)!}{2q^{d-2}}.$$
\end{theorem}

\begin{proof}

It follows directly from Theorem \ref{thm:firstbound} with the choice $s=1$, $r=1$, and Corollary  \ref{cor:MDSq+1} which ensures the existence of the wanted $[n,k,d]_q$-code.

\end{proof}

\section{A lower bound using Almost MDS codes}\label{sec:lowerboundsusingAMDS}

 In Section \ref{sec:lowerusingMDS} we have already studied the bound with respect to MDS codes, hence in this section we will deal with codes with Singleton defect equal to $1$.

\begin{definition}
An $[n,k,d]_q$ code $\C$ with $s(\C)=1$  is called \textbf{Almost MDS} (or \textbf{AMDS} for short).
\end{definition}

Almost MDS codes have been deeply studied in literature, since they represent the closest family to the one of MDS codes.  Some classical examples of those codes arise from algebraic-geometric codes obtained using curves of genus $1$ \cite{ts13}. For the interested reader we refer to \cite{de96, do95, fa97}.

\begin{lemma}\label{lem:AMDSmaxweight}
 Let $q$ be  a prime power, $n,k,d$ be three positive integers such that $d \geq 2$. If $\C$ is an $[n,k,d]_q$ code with $k\leq q-2$, then $\C^\perp$ has a codeword of weight $n$.
\end{lemma}

\begin{proof} Consider a generator matrix for $\C$ that, after permutation of coordinates, we can assume of the form $(I_k \mid A)$.  Then, a generator matrix for $\C^\perp$ is given by $(A^\top \mid -I_{n-k})$.  Since $d\geq2$, the rows of $A$ are all non-identically zero. Indeed, if one of them were identically zero, then we would find a codeword of weight $1$ in $\C$.
Take now an element $c\in \C^\perp$. Then, $c$ is of the form $c=m (A^\top \mid- I_{n-k})$, and we assume $m \in (\Fq^*)^{n-k}$. In this way the last $n-k$ entries of $c$ are non-zero. Therefore, we want to prove that there exists $m\in (\Fq^*)^{n-k}$ such that also the first $k$ entries of $c$ are non-zero. 

Let us call $a_i$ the $i$-th row of $A$, that is also the $i$-th column of $A^\top$. Let us define the sets
$$\mathcal A_i:=\left\{ m \in (\Fq^*)^{n-k} \mid m \in \langle a_i \rangle^\perp\right\}.$$
We want 
$$m \notin \mathcal A:= \bigcup_{i=1}^k \mathcal A_i,$$
so that all the first $k$ entries of $c$ are non zero.  We can give an estimation on the sets $\mathcal A_i$ as follows.
We observe that every $\mathcal A_i$ is described by zeros of a linear polynomial in $n-k$ variables. By Schwartz-Zippel Lemma \cite[Lemma 1]{sc79} we have
$|\mathcal A_i| \leq (q-1)^{n-k-1}$, and hence $ |\mathcal A|\leq k(q-1)^{n-k-1}$. Since $k\leq q-2$, we conclude observing that
$$|(\Fq^*)^{n-k}|=(q-1)^{n-k}> k(q-1)^{n-k-1} \geq |\mathcal A|.$$



\end{proof}

In Section \ref{sec:lowerbounds}, we have introduced the function $M(\mathcal K,d)$ for any positive integer $d$ and any subgroup $\mathcal K$ of some symmetric group. In the special case that $\mathcal K$ is the direct product of copies of $\Z/2\Z$, we can associate the function $M(\mathcal K,d)$ to a very well-known function in  coding theory.

\begin{definition}
Let $q$ be a prime power, and $d,n$ be two positive integers such that $d\leq n$. We define the number
$A_q(n,d)$ 
as the maximum cardinality of a non-necessarily linear code of length $n$ and minimum distance $d$ over $\Fq$, i.e.
$$A_q(n,d)=\max\{ |\C| : \C \subseteq \F_q^n, d(\C)=d\}.$$
\end{definition}

\begin{lemma}\label{lem:KA2}
 Let $\mathcal K \subseteq \mathcal S_{2n}$ be a subgroup of the form $\mathcal K \cong (\mathcal S_{2})^r\cong (\Z/2\Z)^r$. Then $M(\mathcal K,d)=A_2(r,\lfloor \frac{d}{2} \rfloor)$.
\end{lemma}

\begin{proof}
The subgroup $\mathcal K$ can be seen as, after relabeling the elements $\{1,\ldots, 2n\}$, the subgroup
$$\mathcal K=\mathcal S_{\{1,2\}}\times\mathcal S_{\{3,4\}} \times \ldots \times \mathcal S_{\{2r-1,2r\}} =\langle \{(2i-1,2i) \mid i=1,\ldots, r\} \rangle. $$
 The map
$$\begin{array}{rccl}
 \phi :& \F_2^r & \longrightarrow & \mathcal K \\
&v=(v_i)_i & \longmapsto & \prod_i (2i-1,2i)^{v_i}
\end{array}$$
is a bijective homothety, i.e. it preserves the distance up to a scalar multiple. In fact, we have that for every $u,v \in \F_2^r$
$$2d_H(u,v)=d_H(\phi(u),\phi(v)).$$
Therefore $|\phi(A_2(r,\lfloor \frac{d}{2} \rfloor))|\leq |M(\mathcal K,d)|$ by the maximality of $M(\mathcal K,d)$  and $|A_2(r,\lfloor \frac{d}{2} \rfloor)|\geq \phi^{-1}(|M(\mathcal K,d)|)$ by the maximality of $A_2(r,\lfloor \frac{d}{2} \rfloor)$. The claim follows as $\phi$ is a bijection.
\end{proof}

\begin{theorem}\label{thm:AMDSbound}
Let $n,d$ be two positive integers such that $d\leq n$ and $q$ be a prime power with $q<n\leq 2q$. If there exists  an $[n,n-d,d]_q$ AMDS code $\C$ such that $\C^\perp$ has a codeword of weight $n$, then
$$M(n,d) \geq \frac{n! A_2(n-q,\lfloor \frac{d}{2} \rfloor)}{2^{n-q}q^{d-1}}.$$
\end{theorem}

\begin{proof}
 It directly follows from Theorem \ref{thm:firstbound} with $s=1$, $r=n-q$, 
 (and therefore $\mathcal K=\mathcal (S_2)^r$), and Lemma \ref{lem:KA2}.
\end{proof}

\begin{theorem}\label{thm:AMDSbound2}
Let $n,d$ be two positive integers such that $d\geq 2$ and $q$ be a prime power with $q<n\leq \min\{2q, q+d-2\}$. If there exists  an $[n,n-d,d]_q$ AMDS code $\C$, then
$$M(n,d) \geq \frac{n! A_2(n-q,\lfloor \frac{d}{2} \rfloor)}{2^{n-q}q^{d-1}}.$$
\end{theorem}

\begin{proof}
It follows from Theorem \ref{thm:AMDSbound} and Lemma \ref{lem:AMDSmaxweight}.
\end{proof}

\section{Comparison with the previous bounds}\label{sec:comparison}

We explain here how our bounds compare with others given in the literature. As our Theorem \ref{cor:MDSq} allows $q$ to be the next prime power greater or equal to $n$, we beat (or at least equal) the bounds in \cite{ji16,wa17} (see Table \ref{tab:comparison}).
Interestingly enough, when $n-1$ is a prime power, Theorem \ref{cor:MDSq+1} beats asymptotically the bounds in \cite{ji16,wa17} in the large distance regime. We formalize this in the proposition below.
Let us denote by $\mathrm{nextprime}(\cdot)$ the function that sends an integer $n$ to the smallest prime number larger than or equal to $n$, and by  $\mathrm{nextprimepower}(\cdot)$ the function that sends an integer $n$ to the smallest prime number larger than or equal to $n$.

%
%

For the rest of this section, we set 
 \begin{align*}
B_{\text{old}}(n,d) &=\frac{n!}{\mathrm{nextprime}(n)^{d-2}}, \\
B_{\text{mds}}(n,d)&=\frac{n!}{\mathrm{nextprimepower}(n)^{d-2}},\\
B_{\text{new}}(n,d)&=\frac{n!}{2(n-1)^{d-2}}.\\
\end{align*}

More specifically, $B_{\text{old}}(n,d) $ represents the bound in \cite[Theorem 2]{ji16} and \cite[Theorem 13]{wa17}, while $B_{\text{mds}}(n,d)$ and $B_{\text{new}}(n,d)$ are the bounds in Theorem \ref{thm:boundMDSq} and Theorem \ref{thm:boundMDSq+1}, respectively. 
It is trivial to see that  $B_{\text{mds}}(n,d)\geq B_{\text{old}}(n,d)$, for every $n,d$. 
We now focus on the comparisons of $B_{\text{old}}(n,d) $ with $B_{\text{new}}(n,d)$
and the bound given in Theorem \ref{thm:AMDSbound}.

\begin{proposition}\label{prop:comparisonlarged2}
Let $n \in \mathbb N$, and set $d=bn$ for some $0<b<1$.
Then, 
$$\liminf_{n} \frac{B_{\text{new}}(n,d)}{B_{\text{old}}(n,d)}\geq \frac{e^b}{2}.$$
 In particular, for $b>\log_e(2)$,  $B_{\text{new}}(n,d)$ gives asymptotically a better bound than $B_{\text{old}}(n,d)$.
\end{proposition}

\begin{proof}
We have,
\[\frac{B_{\text{new}}(n,d)}{B_{\text{old}}(n,d)}\geq \frac{n^{d-2}}{2(n-1)^{d-2}}=\frac{1}{2}\left(1+\frac{1}{n-1}\right)^{bn-2} \longrightarrow \frac{e^b}{2}.\]
\end{proof}
It is important to show that in the regime where we beat the old bound, the new one is actually non-trivial. We do that in the following remark.
\begin{remark}
Observe that in the regime  $\log_e(2)<\frac{d}{n}<1$, the bound $B_{\text{new}}(n,d)$ is asymptotically non-trivial. Indeed, 
\[B_{\text{new}}(n,d)=\frac{n!}{2(n-1)^{bn-2}}\geq \sqrt{2\pi} \frac{n^{n+\frac{1}{2}}}{2e^{n}(n-1)^{bn-2}}>\sqrt{2\pi}\frac{n^{(1-b)n}}{2e^{n}}\longrightarrow +\infty,\]
where the second inequality follows from Stirling's approximation formula.
Moreover, notice that the bound $B_{\text{new}}(n,d)$ can only be used when $n-1$ is a prime power. 
\end{remark}

The following proposition shows the regime in which our bound in Theorem \ref{thm:AMDSbound2} beats by a large scale the previous known bounds.
\renewcommand{\arraystretch}{1.3}

\begin{table}\begin{center}
\begin{tabular}{|N{2}{0}| N{18}{0}| N{16}{0}| N{18}{0}|}
\hline
{\hfill n} & {\hfill Theorem \ref{thm:boundMDSq}} & {\hfill Theorem \ref{thm:boundMDSq+1}} & {\hfill \cite{ji16,wa17}}  \\[0.1cm]
\hline
9 & \multicolumn{1}{r|}{\textbf{\numprint{56}}} & 45 & 25 \\
\hline
10  & 248 & \multicolumn{1}{r|}{\textbf{\numprint{277}}} & 248 \\
\hline
11 & \multicolumn{1}{r|}{\textbf{\numprint{2727}}} & & \multicolumn{1}{r|}{\textbf{\numprint{2727}}} \\
\hline
12 & \multicolumn{1}{r|}{\textbf{\numprint{16772}}} & 16359 & \multicolumn{1}{r|}{\textbf{\numprint{16772}}} \\
\hline
13 &  \multicolumn{1}{r|}{\textbf{\numprint{218026}}} & & \multicolumn{1}{r|}{\textbf{\numprint{218026}}} \\
\hline
14 &1330236 & \multicolumn{1}{r|}{\textbf{\numprint{1526178}}} & 1043789 \\
\hline
15 & \multicolumn{1}{r|}{\textbf{\numprint{19953528}}} & & 15656834 \\
\hline
16 & \multicolumn{1}{r|}{\textbf{\numprint{319256438}}} &  & 250509332 \\
\hline
17 &  \multicolumn{1}{r|}{\textbf{\numprint{4258658638}}} & 2713679719 & \multicolumn{1}{r|}{\textbf{\numprint{4258658638}}}\\
\hline
18 & \multicolumn{1}{r|}{\textbf{\numprint{49127720826}}} & 38327927742 & \multicolumn{1}{r|}{\textbf{\numprint{49127720826}}} \\
\hline
19 & \multicolumn{1}{r|}{\textbf{\numprint{933426695689}}} & &\multicolumn{1}{r|}{\textbf{\numprint{933426695689}}}  \\
\hline
20 & 8693872621156 & \multicolumn{1}{r|}{\textbf{\numprint{9334266956886}}} &8693872621156 \\
\hline
21 & \multicolumn{1}{r|}{\textbf{\numprint{182571325044256}}} & & \multicolumn{1}{r|}{\textbf{\numprint{182571325044256}}} \\
\hline
\end{tabular}
\caption{This table compares the results given by Theorem \ref{thm:boundMDSq} and Theorem \ref{thm:boundMDSq+1} with \cite[Theorem 2]{ji16} and \cite[Theorem 13]{wa17} for many values of $n$ and $d=6$. For each line of the table the numbers in bold denote the best bounds.}\label{tab:comparison}
\end{center}
\end{table}

\begin{proposition}\label{prop:comparisoncertaind}
Let $q$ be a prime power and $n=\alpha q$ for some $\alpha$  such that  $1<\alpha \leq 2$.  Set $d=bn=b\alpha q$  with $\frac{\alpha-1}{\alpha \log_2(\alpha)}<b<1$, and 
\[B_{\mathrm{amds}}(n,d)=\frac{n!A_2(n-q,\lfloor \frac{d}{2}\rfloor)}{2^{n-q}q^{d-1}}.\]
Then 
$$\frac{B_{\mathrm{amds}}(n,d)}{B_{\mathrm{old}}(n,d)}\longrightarrow +\infty,$$
as $n$ goes to infinity.
\end{proposition}
\begin{proof}
We have
$$ \frac{B_{\text{amds}}(n,d)}{B_{\text{old}}(n,d)}\geq \frac{A_2((\alpha-1)q,\lfloor \frac{b\alpha q}{2}\rfloor)\alpha^{b\alpha q-2}q^{b\alpha q-2}}{2^{(\alpha-1)q}q^{b\alpha q -1}}\geq \frac{2}{\alpha^2 q}\frac{\alpha^{b\alpha q}}{2^{(\alpha-1)q}}
=\frac{2}{\alpha^2 q}2^{(\alpha\log_2(\alpha) b-\alpha+1)q}.$$
Since we assumed $\frac{\alpha-1}{\alpha \log_2(\alpha)}<b$, then $\alpha \log_2(\alpha)b-\alpha+1>0$, and in turn $ \frac{B_{\text{amds}}(n,d)}{B_{\text{old}}(n,d)}\longrightarrow +\infty$. 
\end{proof}
Again, we notice in the next remark that in the regime where we beat the old bound, the new one is actually non-trivial.
\begin{remark}
Observe that in the regime $n=\alpha q$ and $d=b n=b\alpha q$, with $b<1$ the bound $B_{\text{amds}}(n,d)$ is non-trivial. Indeed
\[B_{\text{amds}}(n,d)\geq \frac{n!2}{2^{(\alpha-1)q}q^{b\alpha q-1}}\geq 2\sqrt{2\pi}q^{\frac{3}{2}}\frac{\alpha^{\alpha q+\frac{1}{2}}q^{(1-b)\alpha q}}{e^{\alpha q}2^{(\alpha-1)q}}\longrightarrow +\infty, \]
for $q$ going to infinity,  where the second inequality follows from Stirling's formula.
\end{remark}

\begin{remark}
Proposition \ref{prop:comparisoncertaind} shows that the bound given in Theorem \ref{thm:AMDSbound2} could beat by far the bound in \cite[Theorem 2]{ji16} and \cite[Theorem 13]{wa17}, and therefore also the one from Theorem \ref{thm:boundMDSq}, for $\frac{d}{n}>\frac{\alpha-1}{\alpha \log_2(\alpha)}$ and $n$ large enough. The reader should notice that Proposition \ref{prop:comparisoncertaind} is conditioned to the existence of a family of AMDS codes of length $n$ over $\Fq$, for $q$ large and $1<\frac{n}{q} \leq 2$ fixed. The existence of such family is not proven nor disproven and explicit constructions of linear codes with these parameters becomes now central also in the theory of permutation codes.
%
%
\end{remark}

\section{Conclusions}\label{sec:conclusions}

In this paper we connected the theory of linear codes with the theory of permutation codes.
In turn, this allows to produce new lower bounds for the maximal size of permutation codes. The lower bounds produced use the existence of certain codes of given distance and length over an alphabet of a given size, converting the problem of finding a lower bound for permutation codes of given distance into the problem of finding a certain linear codes with parameters as in Theorem \ref{thm:firstbound}.
In Section \ref{sec:comparison} we apply Theorem \ref{thm:firstbound} and obtain improved bounds with respect to the ones in the literature \cite{ji16,wa17}, as one can now select a the next prime power instead of the next prime in the bound of \cite[Theorem 2]{ji16} and 
\cite[Theorem 13]{wa17}  (thanks to our Theorem \ref{thm:boundMDSq}). Moreover, in Proposition \ref{prop:comparisonlarged2} and Proposition \ref{prop:comparisoncertaind} we show that we beat them asymptotically for certain regimes of $n$ and $d$.

\bibliographystyle{abbrv}
\bibliography{biblio}

\end{document}